\documentclass[10pt,twocolumn,letterpaper]{article}

 \usepackage[pagenumbers]{cvpr}
\usepackage[dvipsnames]{xcolor}

\newcommand{\agt}{${\text{GT}}^*$}


\usepackage[dvipsnames]{xcolor}
\definecolor{projectpink}{HTML}{E91E63} 

\usepackage[colorlinks]{hyperref}
\hypersetup{
  linkcolor=black,
  citecolor=black,
  urlcolor=projectpink,
}
\usepackage[T1]{fontenc}
\usepackage[colorlinks]{hyperref}
\usepackage{mathrsfs}
\usepackage{wrapfig}
\usepackage{caption} 
\usepackage{capt-of}
\usepackage[dvipsnames]{xcolor}
\usepackage{multirow}
\usepackage{booktabs}

\definecolor{cvprblue}{rgb}{0.21,0.49,0.74}

\title{
  CaricatureGS: Exaggerating 3D Gaussian Splatting \\[2pt]
  Faces with Gaussian Curvature
}

\author{
Eldad Matmon\quad
Amit Bracha\quad
Noam Rotstein\quad
Ron Kimmel\\ [10pt]
Technion -- Israel Institute of Technology, Haifa, Israel\\[1pt]
}

\begin{document}
\twocolumn[{%
\maketitle
\begin{center}
  \includegraphics[width=.8\linewidth]{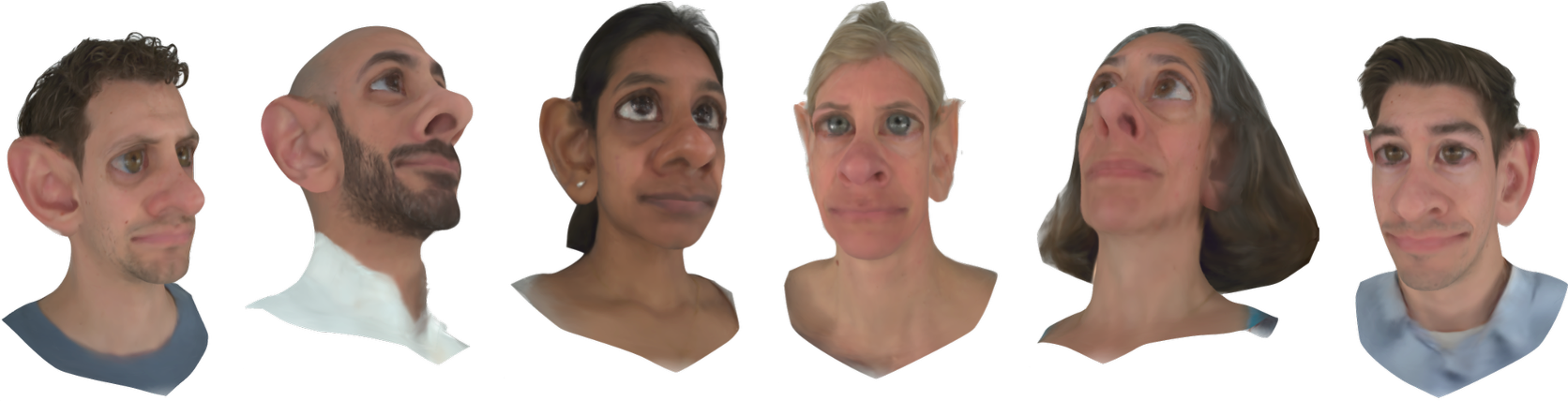}
  \captionof{figure}{Photorealistic 3D caricature avatars produced by our method.}
  \label{fig:teaser}
\end{center}
}]
\begin{abstract}
A photorealistic and controllable 3D caricaturization framework for faces is introduced.
We start with an intrinsic Gaussian curvature-based surface exaggeration technique, which, when coupled with texture, tends to produce over-smoothed renders.
To address this, we resort to 3D Gaussian Splatting (3DGS), which has recently been shown to produce realistic free-viewpoint avatars.
Given a multiview sequence, we extract a FLAME mesh, solve a curvature-weighted Poisson equation, and obtain its exaggerated form.
However, directly deforming the Gaussians yields poor results, necessitating the synthesis of pseudo–ground-truth caricature images by warping each frame to its exaggerated 2D representation using local affine transformations.
We then devise a training scheme that alternates real and synthesized supervision, enabling a single Gaussian collection to represent both natural and exaggerated avatars.
This scheme improves fidelity, supports local edits, and allows continuous control over the intensity of the caricature. 
In order to achieve real-time deformations, an efficient interpolation between the original and exaggerated surfaces is introduced.
We further analyze and show that it has a bounded deviation from closed-form solutions.
In both quantitative and qualitative evaluations, our results outperform prior work, delivering photorealistic, geometry-controlled caricature avatars.

\noindent Project page:
\href{https://c4ricaturegs.github.io/}{\textcolor{projectpink}{https://c4ricaturegs.github.io}}

\end{abstract}


\section{Introduction}\label{sec:intro}

Face caricaturization refers to the action of exaggerating distinctive facial features while preserving identity.
Despite its promise for lifelike, immersive avatars, producing such exaggerations in controllable, photorealistic 3D remains an open challenge.
Successful mesh-based approaches are based on geometric deformations with curvature-based methods, such as the scale-aware Poisson framework  \cite{sela_computational_2015-1}.
When such deformed surfaces are rendered through traditional mesh-centric pipelines, such as texture mapping, the results often appear unnatural \cite{sela_computational_2015-1}. 
Recently, 3D Gaussian Splatting (3DGS) \cite{kerbl_3d_2023} has emerged as a potential multiview representation that provides state-of-the-art real-time photorealism by optimizing Gaussian primitives directly from a given set of images taken from various directions.

This raises the following question.

\emph{
Can we combine curvature-based geometric fidelity with 3DGS to generate photorealistic caricatures?
}

To address this, we start with a multiview video of a subject and its extracted FLAME mesh \cite{FLAME:SiggraphAsia2017}.
From this, solving the weighted Poisson equation gives us the deformed caricature mesh.
We rig Gaussians to the original undeformed surface and train them following a framework previously proposed for facial expressions \cite{lee_surfhead_2024}. 
Later, at inference, we deform the original mesh and its rigged Gaussians according to the caricature mesh, stretching, shearing, and rotating them.
However, modeling these deformations as merely an additional expression, using Gaussians optimized only on the input sequence, leads to low fidelity (see~\cref{fig:Rendering_Results}), revealing a domain gap in which caricatures lie outside the distribution of natural expression dynamics.

To bridge this gap and in the absence of real caricature training data, we synthesize pseudo–ground truth (\agt) by warping each input frame with \emph{Local Affine Transformations} (LAT) induced by the correspondence from the original mesh to its curvature-exaggerated counterpart, producing photorealistic supervision (see \cref{subsec:LAT_GT}). 
During training, we stochastically alternate between real views and \agt views so that a single Gaussian set jointly models both natural and caricatured deformations, allowing the Gaussians to benefit from real ground truth while adapting to \agt. 
To mitigate occlusion-related artifacts and protect fine structures (\eg  hair and mesh boundaries), we apply a spatial mask that freezes the affected Gaussians during \agt steps (\cref{fig:hair_error}). 
These Gaussians are updated only from real frames, allowing a consistent appearance to accumulate in their attributes.

Although trained only on the two sets of views, the optimized model offers additional flexibility and control at inference.
First, it generalizes across a continuous range of caricature intensities, with the exaggeration level controlled by an efficient linear interpolation as an approximation of the solution to the weighted Poisson equation, a property that we demonstrate both theoretically and empirically.
Moreover, this representation is robust to both global and local deformations, enabling controlled localized edits, such as exaggerating the nose size, while leaving unrelated regions unchanged.

The new 3DGS animatable representation is the first, to our knowledge, to enable photorealistic caricature rendering while faithfully retaining identity under caricature deformations.
We compare it to the current state-of-the-art dynamic facial reconstruction model \cite{lee_surfhead_2024}, which consistently achieves higher scores and qualitative results in terms of image fidelity, structural consistency, and identity preservation metrics.

\noindent
\textbf{Our contributions include,}
\begin{itemize}
\item A novel 3DGS training scheme that uses \agt generated with local Affine transformations that represent real and caricature avatars.
\item Curvature-weighted deformation with rigged 3DGS for identity-preserving photorealistic caricatures.
\item Real-time avatars supporting variable exaggeration levels and fine-grained local control of facial features.
\end{itemize}

\section{Related Work}
\label{sec:related}
\subsection{Representation for 3D Head Avatars}

Neural implicit representations have become a dominant approach for high-fidelity 3D head avatars, enabling photorealistic view synthesis from sparse multiview observations.

IMAvatar~\cite{IMAVATAR} combines 3D morphable-model parameters for pose and expression control using neural blendshapes and skinning fields to produce animatable head avatars. 
ImFace~\cite{IMFACE} disentangles identity and expression using two deformation fields applied to a signed distance function (SDF) template.
ImFace++~\cite{IMFACE++} extends this approach with a two-stage refinement framework that improves detail preservation.

NeRFs~\cite{NeRF} map spatial coordinates and viewing directions to radiance and density and render images via volumetric integration. 
For head avatars, Wang et al.~\cite{Learning_Compositional_Radiance_Fields} encode sparse views into a 3D structure-aware grid of animation codes refined by an MLP. Gafni et al.~\cite{Gafni_Dynamic} integrate a low-dimensional morphable face model with a neural scene representation to obtain photorealistic, controllable avatars from monocular video. 
Gao et al.~\cite{Reconstructing_Personalized_Semantic_Facial_NeRF} employ multilevel voxel fields with low-dimensional expression coefficients to capture elements beyond mesh blendshapes (\eg hair and accessories). 
INSTA~\cite{INSTA} accelerates dynamic NeRF by embedding it around a surface representation to obtain animatable avatars from short monocular video and AvatarMAV~\cite{AvatarMAV} decouples appearance from motion via motion-aware neural voxel grids.

3D Gaussian splatting ~\cite{kerbl_3d_2023} represents 3D scenes as anisotropic Gaussian primitives, and renders them via differentiable splatting. 
In the context of head avatars, Rig3DGS~\cite{rig3dgs} reconstructed scenes in a canonical Gaussian space and learned 3DMM-guided deformations for efficient and photorealistic animation, while HeadGaS~\cite{HeadGas} extended the representation with blendable Gaussians whose attributes adapt to expression coefficients. 
MeGA~\cite{MeGA} introduced a hybrid mesh–Gaussian design, combining splats with mesh geometry for high-fidelity rendering and editable head avatars. GaussianAvatars~\cite{qian_gaussianavatars_2024} bound deformable 3D Gaussians to a parametric face mesh via a binding inheritance strategy, and SurFhead~\cite{lee_surfhead_2024} replaced the 3D Gaussians with 2D Gaussian surfels \cite{huang_2d_2024}, applying Jacobian Blend Skinning and polar decomposition, achieving state-of-the-art results in dynamic head reconstruction.

\begin{figure*}[t]
  \centering
  \makebox[\textwidth][c]{%
    \includegraphics[width=1.06\textwidth]{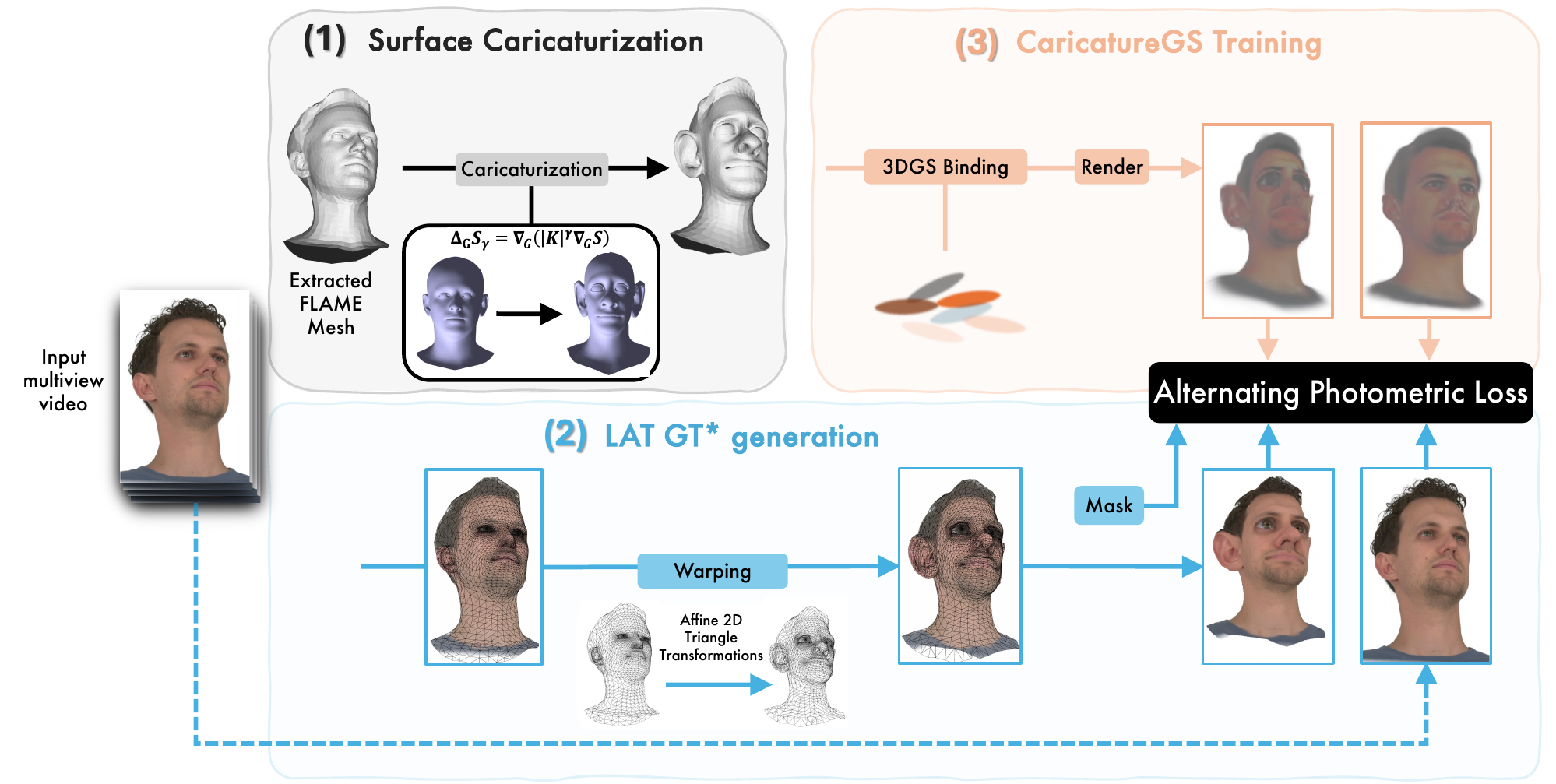}%
  }
  \caption{
  \textbf{CaricatureGS generation framework.}
    \textbf{(1)}\label{fig:pipeline_step1} From a subject’s multi-view video, we extract a FLAME mesh and compute a curvature-driven caricature based on it. 
    Combined with subject-specific FLAME parameters, this yields the subject’s caricature mesh.
    \textbf{\textcolor{blue}{(2)}}\label{fig:pipeline_step2} Per-triangle 2D affine transforms map the neutral mesh projection to its caricatured counterpart, warping each frame to generate pseudo–ground-truth image pairs.
    \textbf{\textcolor{orange}{(3)}}\label{fig:pipeline_step3} Anisotropic 3D Gaussians primitives are bound to the original mesh and transformed to the caricature mesh via the corresponding 3D triangle transforms. 
    Rendered neutral and caricature views are alternated and compared to their pseudo–ground-truth counterparts in joint optimization.
    }
    \vspace{-6pt}
  \label{fig:pipeline}
\end{figure*}

\subsection{Mesh Deformation and Exaggeration}\label{sec:mesh_deform_ex}
Classical mesh-based approaches realize deformations using geometry processing, e.g., Poisson/Laplacian editing and related curvature-driven deformations \cite{ARAP_modeling_2007, LaplacianMeshEditing_2004, GeoFilter, yu2004mesh}.
For faces, mesh-based deformation and caricaturization have been explored through both geometry-driven and data-driven approaches, evolving from early parametric face models to modern neural deformation networks. 
Early work by Blanz and Vetter \cite{blanz_morphable_nodate} introduced the 3D Morphable Face Model (3DMM), representing shape and texture as linear combinations of example faces, enabling identity and expression manipulation. 
In the caricature domain, Brennan \cite{brennan_caricature_1985} developed an interactive system for producing line-drawn caricatures by exaggerating the vector differences between the features of a subject and an average face. 
Eigensatz \cite{eigensatz_curvature-domain_2008} used curvature maps to enhance, smooth, and transfer characteristics while preserving global structure. 
Later, Sela et al. \cite{sela_computational_2015-1} proposed a scale-aware Poisson-based curvature framework for surface caricaturization, exaggerating geometric features while maintaining spatial and temporal coherence. 

Data-driven methods have enabled for more expressive and automated mesh exaggerations. 
Wu et al.~\cite{wu_alive_2018} learned deformation patterns from artist-created examples to generate 3D caricatures from a single 2D portrait while preserving identity. 
Han et al.~\cite{han_deepsketch2face_2017} introduced \emph{DeepSketch2Face}, where a CNN infers and refines 3D face or caricature meshes from 2D sketches, while their later work \emph{CaricatureShop}~\cite{han_caricatureshop_2018} combined vertex-wise Laplacian scaling with deep learning to produce photorealistic, personalized 2D caricatures from reconstructed 3D faces. 
Jung et al.~\cite{jung_deep_2022} advanced this idea by using an MLP to map latent codes to 3D displacements, supporting controlled and diverse exaggerations. 
More recent approaches focus on style adaptation and broader correspondences. 
Yan et al.~\cite{yan_cross-species_2022} presented an alignment-aware 3D face morphing framework with controller-based mapping for cross-species correspondence. 
Olivier et al.~\cite{olivier:hal-03763591} explored GAN-based style transfer from scans to caricatures. Yoon et al.~\cite{yoon_lego_2024} proposed \emph{LeGO}, a one-shot method that fine-tunes a surface deformation network to replicate a target style.
An additional line of work that can be adapted to facial exaggeration is the generative line, exemplified by Diffusion- and GAN-based 3DGS editors \cite{wang_gaussianeditor_2024, chen_gaussianeditor_2024, li_generating_2024}, which operate primarily on appearance while leaving the underlying geometry unchanged.

\section{Method}
\label{sec:method}

Here, we introduce a method for creating controllable photorealistic caricaturizations of human faces with 3DGS.  
Our pipeline, illustrated in~\cref{fig:pipeline}, begins with a multiview video of a subject, from which we extract a FLAME-fitted mesh.  
In~\cref{subsec:Surface_Caricaturization}, we describe how we deform the geometry to obtain a caricaturized mesh.
To supervise 3DGS training, we generate pseudo–ground-truth caricature images (\agt) using a 2D warping scheme (\cref{subsec:LAT_GT}).  
The Gaussian primitives are then rigged to both the neutral and caricatured meshes and optimized by minimizing alternating photometric losses between their renders, the original frames, and the corresponding \agt{} images (\cref{subsec:caricatureGS_training}).
Finally, we demonstrate that this single shared Gaussian set, although trained only on these two image domains, supports real-time rendering across a continuous range of exaggeration levels via surface interpolation and enables region-specific edits (\cref{subsec:Interpolation}).

\subsection{Surface Caricaturization}
\label{subsec:Surface_Caricaturization}
Starting from the temporally consistent FLAME mesh obtained by fitting the landmarks~\cite{face2face}, we apply a curvature-driven deformation that exaggerates facial geometry.
Since the mesh maintains consistent vertex correspondences across frames, these deformations preserve temporal coherence.
To implement this deformation, we formulate it as a weighted Poisson equation on the surface. 

Let \(S\in \mathbb{R}^3\) be a surface with metric \(G\) and Gaussian curvature \(K(p)\) for \(p\in S\). 
For \(\gamma\in[0,\gamma_f]\), we define the \emph{weighted Poisson equation} 
\begin{eqnarray}
\Delta_G S_\gamma &=&
\nabla_G\!\cdot\!\big(w(\gamma)\nabla_G S\big).
\label{general_weighted_poisson}
\end{eqnarray}
We adopt the curvature-driven deformation model introduced by~\cite{sela_computational_2015}, whose weights are given by \( w(\gamma) = |K|^{\gamma} \). This gives, for each \(\gamma\), the following family of Poisson equations :
\begin{eqnarray}
\Delta_G S_\gamma &=&
\nabla_G\!\cdot\!\big(|K|^{\gamma}\nabla_G S\big).
\label{curvature_weighted_poisson}
\end{eqnarray}
In order to derive the deformed surface we solve the PDE by the following least-squares:
\begin{eqnarray}
\min_{\tilde{x}} \| L \tilde{x} - b \|^2_A.
\end{eqnarray}
\(L\) is the \emph{discrete Laplace–Beltrami operator}, defined as \(L = A^{-1} W\), \( A \) is a diagonal area matrix, \( W \) is the classic \emph{cotangent weight matrix} and \(b = \nabla_G \cdot \bigl(|K|^\gamma \nabla_G(x)\bigr)\).
The weighted norm is defined as \(\|F\|_A^2 = \operatorname{trace}(F^T A F)\).
We denote by \(S_\gamma\) the solution of the weighted Poisson equation in equation \ref{curvature_weighted_poisson}.

To accommodate open surfaces, where the Gaussian curvature may be ill defined on \(\partial S\) or to allow precise user-controlled exaggerations as discussed in \cref{subsec:Interpolation}, we impose boundary conditions on the selected vertices, namely:
\begin{equation}
\min_{\tilde{{x}}\in\mathbb{R}^n}\; \|L\tilde{{x}}-{b}\|_{A}^{2}
\quad \text{s.t.} \quad
B\tilde{{x}}={x}^*,
\label{eq:constrained-ls}
\end{equation}
where \(B\in\{0,1\}^{m\times n}\) selects the rows corresponding to the set of vertices and \({x}^*\) are the prescribed boundary positions. The same constrained system is solved independently for the \(y\) and \(z\) coordinates.

An example of the resulting mesh deformation is illustrated in part~\textbf{(1)} of \cref{fig:pipeline}.


\subsection{
\textbf{\agt} Generation via Local Affine Transforms
}
\label{subsec:LAT_GT}
With these deformed surfaces, the avatar’s geometry is represented in caricatured form.  
For photorealistic rendering, we employ mesh-rigged 3DGS, detailed in \cref{subsec:caricatureGS_training}.
Since using 3DGS without caricature optimization yields poor results (\cref{subsec:baseline}), training requires ground-truth supervision images.  
As real caricature images do not exist, we generate pseudo–ground truth (\agt): photorealistic caricature images that preserve identity while ensuring multiview consistency.

One possible way to obtain such supervision is one-shot stylization (e.g., Zhou et al.~\cite{zhou_deformable_2024}), which narrows the natural–caricature gap using a single exemplar image.  
However, it fails to disentangle style from pose and identity, often transferring both instead of style alone (see supplementary).
We therefore propose an alternative: Local Affine Transformations (LAT), illustrated in part~\textbf{(2)} of \cref{fig:pipeline}.

LAT exploits the shared connectivity of the neutral and deformed meshes, implying a per-triangle correspondence.
Consider corresponding 3D triangles \(X=\{X_1,X_2,X_3\}\in\mathbb{R}^3\) and \(Y=\{Y_1,Y_2,Y_3\}\in\mathbb{R}^3\).  
Let \(\pi:\mathbb{R}^3\to\mathbb{R}^2\) denote the image-plane projection, with \(x_i=\pi(X_i)\) and \(y_i=\pi(Y_i)\in\mathbb{R}^2\).
Assuming \(\{x_1,x_2,x_3\}\) are non-collinear, there exists a unique affine map,
\begin{equation}
\Phi(\mathbf{x})=A\mathbf{x}+\mathbf{b},
\qquad
A\in\mathbb{R}^{2\times2},\;\mathbf{b}\in\mathbb{R}^2,
\end{equation}
such that \(\Phi(x)=y\).
We then used these per-triangle 2D affine transformations to map color from the original image to the 2D projection of the deformed mesh. 
In practice, we apply an inverse warp from each target pixel back to the original image and use bilinear interpolation to avoid empty regions.

Caricature deformation can reveal regions previously self-occluded in the neutral pose or occlude regions that were visible, leaving some pixels in \agt without valid correspondences.  
To address this, we generate 2D triangle-level mask for occluded regions.
In addition, because hair strays fall outside the mesh limits and cannot be warped reliably, we add the hair boundary to the mask.
The final output is pseudo–ground truth (\agt): high-quality caricature images that preserve identity, ensure multiview consistency, and provide effective supervision for 3DGS, together with masks indicating per-pixel validity (see appendix for further details).

\subsection{CaricatureGS Training}
\label{subsec:caricatureGS_training}
We model the avatar’s appearance photorealistically using the 3D Gaussian Splatting framework~\cite{kerbl_3d_2023}.
Each Gaussian \(g_i\) stores local attributes: position \(\mu_i\), scale \(s_i\), rotation \(r_i\), opacity \(\sigma_i\), and a view-dependent color \(c_i\).
At each time frame $k \in [0, N]$, the FLAME mesh $\mathcal{M} \subset \mathbb{R}^3$ is represented by triangles $\{T_j[k]\}_{j=1}^M$, where $M$ is the number of mesh faces.
To ensure spatial–temporal coherence, each Gaussian $G_i$ is linked \cite{qian_gaussianavatars_2024} to a specific triangle $T_j$ by a binding index $b_i$, converting its local attributes to world space.

Building on this rigged Gaussian setup, SurFhead~\cite{lee_surfhead_2024} used 2D Gaussian surfels~\cite{huang_2d_2024}, which represent surfaces as oriented planar Gaussian disks, and replaced Linear Blend Skinning (LBS) with Jacobian Blend Skinning (JBS) for Gaussians deformations, namely,
\begin{eqnarray}
  \Sigma^{1/2}_i &=& \mathbf{J_b}r_i s_i,\quad \mu'_i=\mathbf{J_b}\mu_i + T^x_j \cr
    \text{where }\, 
    \mathbf{J_b} &=& \exp\!\left(\sum_{i\in adj} v_i \log(U_i)\right) \cdot \sum_{i\in adj} v_i P_i, \,\,
\end{eqnarray}
where \(v_i\) are learned weights and \(T^x_j\) is the triangle's barycentric center.
\(U_i\) and \(P_i\) are the rotations and stretches from decomposing the Jacobian gradient \(\mathbf{J}\) via polar decomposition. 
Polar decomposition separates rotation and stretch, ensuring geometrically accurate Gaussian deformations (see~\cite{lee_surfhead_2024} for further details).

We show that a setup originally designed for natural facial expressions can be adapted to caricature modeling by applying the deformed caricature mesh for Gaussian deformation and using \agt for 3DGS optimization.
Nevertheless, training exclusively on \agt introduces occlusion-induced artifacts and limits the model to a single expression level. 
To overcome these limitations, we propose a joint optimization procedure that alternates supervision randomly between real video frames and their caricatured \agt counterparts, while maintaining a single shared set of Gaussians, whose rigging ensures consistent kinematics across both supervision domains.
The masks introduced in \cref{subsec:LAT_GT} prevent supervision of Gaussians corresponding to caricature \agt pixels that cannot be reliably warped. The joint optimization scheme allows the caricatured 3DGS to learn beyond \agt by simultaneously filling occlusion-induced holes using supervision from the original frames.
As further demonstrated in \cref{subsec:hair_mask}, this strategy effectively captures hair details for our caricature avatar, despite hair pixels being excluded from direct \agt supervision.
Moreover, as explained in \cref{subsec:Interpolation}, it also enables the generation of intermediate caricatures at \emph{any} level, at inference, without additional capture.

\begin{figure}[t]
  \centering
    \centering
    \makebox[\linewidth][c]{%
    \includegraphics[width=1.12\linewidth]{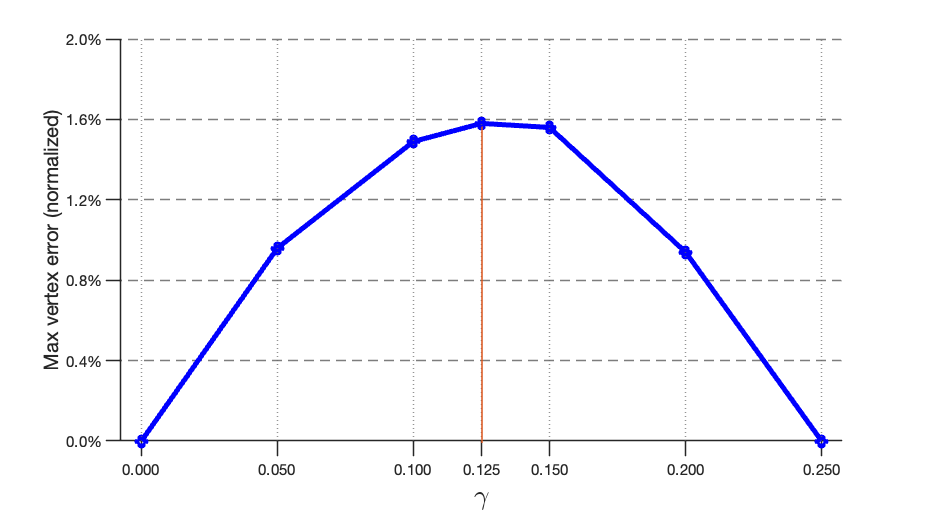}%
    }
    \label{fig:mesh_error}
    \vspace{-16pt}
  \caption{Parametric trend of the error with respect to \(\gamma\).
  The error, normalized by the bounding-box diagonal of the mesh, increases from both ends of \(\gamma\), reaching a negligible maximum at \(\tfrac{\gamma_f}{2}\), where \(\gamma_f=0.25\).
  }
  \vspace{-8pt}
  \label{fig:error_pair}
\end{figure}

\subsection{
CaricatureGS Features
}
\label{subsec:Interpolation}
\vspace{-4pt}
The joint optimization not only complements the caricature Gaussians with information absent from \agt but present in the original frames, it also provides controllability advantages during inference.

\vspace{8pt}
\noindent
\textbf{Controlling Caricature Level}.
After joint training at the target exaggeration level \(\gamma_f\), we empirically observe that the single-rigged Gaussian set generalizes seamlessly, rendering avatars from meshes deformed for any \(\gamma \in [0,\gamma_f]\) without additional optimization.
However, obtaining the deformed mesh for each \(\gamma\) requires solving a curvature-weighted Poisson problem, which poses a runtime bottleneck and makes interactive control of caricature levels impractical.
This motivates the need for a representation that can be efficiently derived from the original mesh \(S_{0}\) and the precomputed caricatured mesh \(S_{\gamma_f}\).
We define this representation as a vertex-wise blend:
\begin{eqnarray}
S_{\mathrm{blend}}(\gamma) &=& (1-\alpha)\,S_0 \;+\; \alpha\,S_{\gamma_f}, \quad \alpha\equiv\frac{\gamma}{\gamma_f}. \label{eq:interpolation_equation}
\end{eqnarray}
We define the residual between the approximation \(S_{\mathrm{blend}}(\gamma)\) and the exact solution \(S(\gamma)\) as
\begin{equation}
\delta S(\gamma) \;=\; S_{\mathrm{blend}}(\gamma)\;-\;S(\gamma).
\end{equation}
In the supplementary material, we show that the \(L^2\) energy of this residual can be bounded using Poincar\'e inequality together with the Lax-Milgram theorem given by
\begin{eqnarray}
\label{eq:secant-bound}
\|\delta S(\gamma)\|_{L^2}
&\lesssim &
\tilde{C}\,\gamma(\gamma_f-\gamma)\,\|\nabla_G S_0\|_{L^2}, \cr
&\, & \cr
\tilde{C}&=& C_P\,(\ln|K|)^2\,e^{\max\{0,\gamma_f\ln|K|\}},
\end{eqnarray}
with \(C_P\) a constant. 

\begin{figure*}[htbp]
  \centering
  \includegraphics[width=1\textwidth]{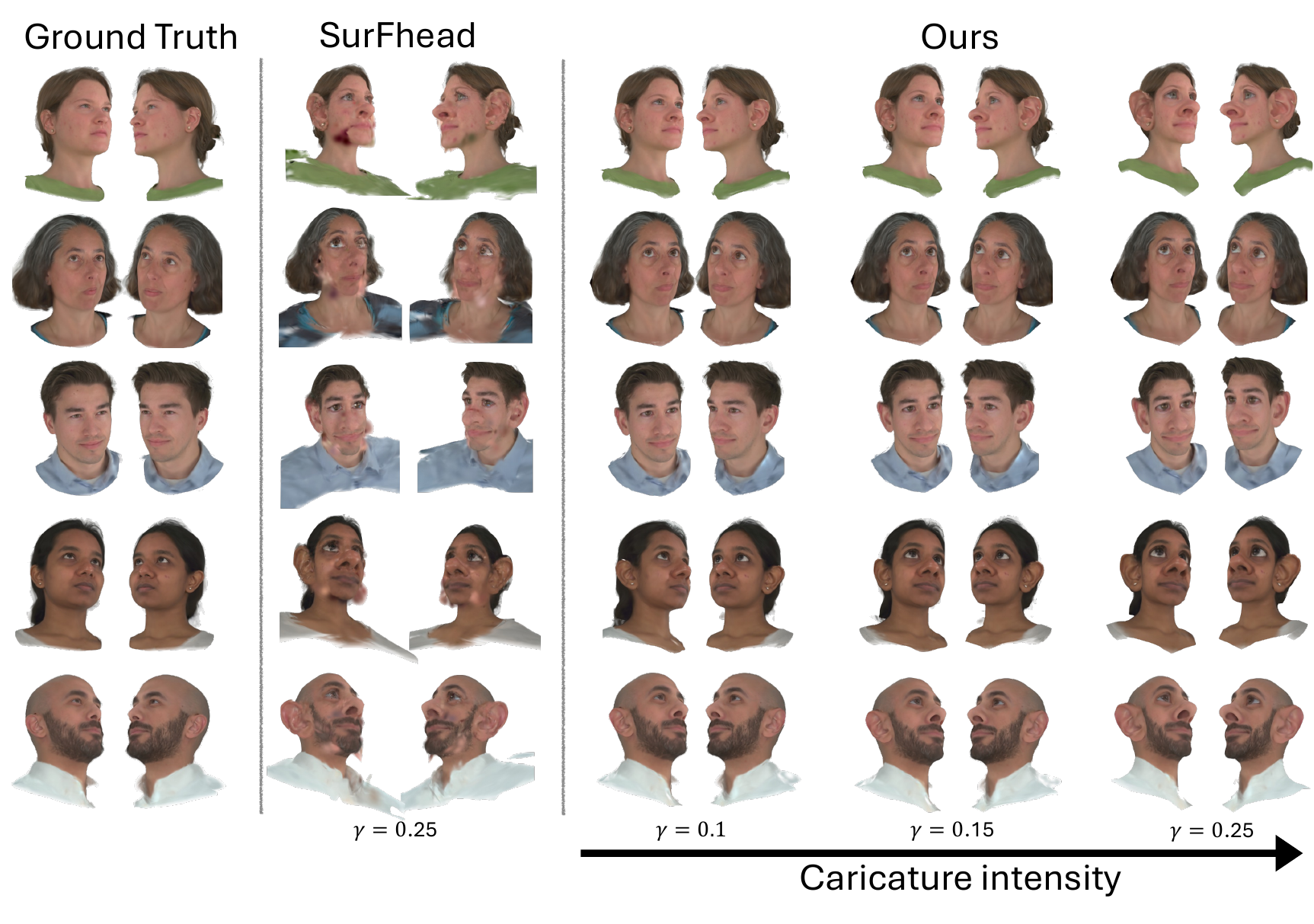}
  \vspace{-24pt}
  \caption{Rendering results from our pipeline \cite{lee_surfhead_2024}.
    \textbf{SURFHEAD:} Caricature generation by first reconstructing an avatar with the state-of-the-art SURFHEAD model~\cite{lee_surfhead_2024}, followed by mesh exaggeration.
  \textbf{Ours:} Renderings across different caricature intensities. 
  Our approximation-based control interpolates smoothly along the caricature intensity axis while preserving visual fidelity.
  }
  \vspace{-10pt}
  \label{fig:Rendering_Results}
\end{figure*}

This bound is zero at the end points \(\gamma=0,\gamma_f\), which means there is no error, as expected from \eqref{eq:interpolation_equation} and maximized near \(\gamma=\frac{\gamma_f}{2}\), where it remains small in practice. 
Empirically, we evaluate the maximal deformation error between \(S_{\mathrm{blend}}(\gamma)\) and \(S_\gamma\) on varying \(\gamma\) and different subjects, normalized by the mesh bounding-box diagonal.
As shown in \cref{fig:error_pair}, the worst-case deviation is negligible, supporting the fidelity of the interpolation and confirming that it lies near the theoretical midpoint of the exaggeration, as predicted. 
This implies that, with this approximation, no additional Poisson equations need to be solved when inferring new \(\gamma\) values, thereby enabling full interactive control of caricature levels.
In \cref{fig:Rendering_Results}, we illustrate that this interpolation scheme enables a single set of Gaussians to smoothly represent shape deformations across the full range of \(\gamma\).

\vspace{8pt}
\noindent
\textbf{Localized Caricature Control.}
Our curvature-weighted model uses the local curvature $K$ to generate a globally consistent caricature by solving the unconstrained Poisson equation. To target specific regions, we solve the constrained least-squares system in \cref{eq:constrained-ls}, whereby only the chosen region of interest undergoes curvature deformations, producing a smooth and localized exaggerations that blend harmonically with the rest of the face. 
Coupled with the training scheme in \cref{subsec:caricatureGS_training}, the 3DGS,  rigged to the mesh, faithfully tracks these deformations, so the same Gaussian set realizes semantically controlled exaggerations while preserving identity and global shape (see \cref{fig:semantic_rendering}).

\begin{figure}[t]
  \centering
    \includegraphics[width=0.85\linewidth]{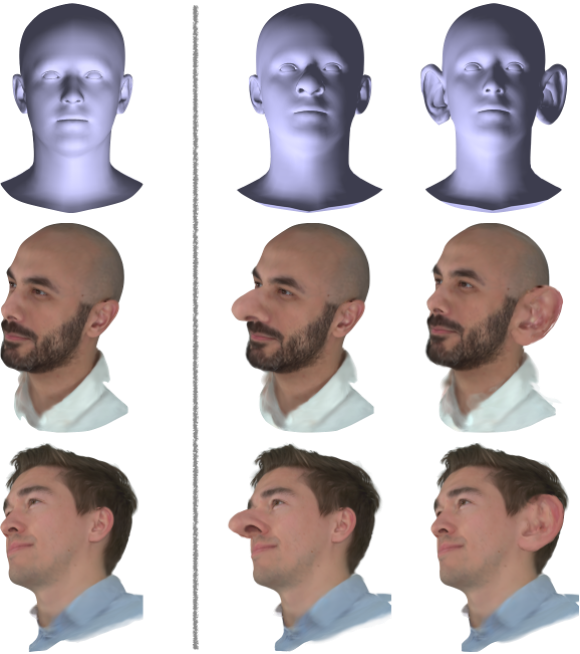}
    \label{fig:semantic_rendering}
    \caption{Visualizations of localized, semantically controlled facial exaggerations.}
    \vspace{-14pt}
  \label{fig:semantic_rendering}
\end{figure}

\vspace{-3pt}
\section{Experiments}
\label{sec:experiments}
\vspace{-3pt}

We evaluate our caricaturized avatars along two main axes:
(i) photorealistic rendering,
(ii) identity preservation.
All experiments are conducted on the NeRSemble dataset~\cite{kirschstein_nersemble_2023} and compared against the recent state-of-the-art 4D avatar reconstruction method of SurFhead ~\cite{lee_surfhead_2024}. 
Unless noted otherwise, we apply an unconstrained exaggeration with $\gamma_f=0.25$. 

\vspace{-3pt}
\subsection{Dataset}
\label{subsec:dataset}
\vspace{-3pt}
The NeRSemble dataset~\cite{kirschstein_nersemble_2023} provides a multi-view facial performance dataset captured by $16$ spatially arranged, synchronized high-resolution cameras. 
It comprises $10$ scripted sequences, $4$ emotion-driven (EMO) and $6$ expression-driven (EXP), plus an additional free self-reenactment sequence. 
For fair comparison, we adopt the same train/validation/test partition as in~\cite{lee_surfhead_2024} with $120,000$ training iterations. 
Further implementation details are provided in the supplementary.

\subsection{Baseline}
\label{subsec:baseline}

To the best of our knowledge, there are no explicit methods that construct a dynamic 3D photorealistic model from an input multi-view video.
To this end, we compare with SurFhead~\cite{lee_surfhead_2024} using the authors’ official implementation.
SurFhead achieves state-of-the-art performance in head reconstruction and reenactment and, in principle, can handle mesh deformations through JBS, making it the most suitable baseline for comparison.
We train the SurFhead on the original input sequence and, at inference, we exaggerate the underlying mesh using \(\gamma_f\), as elaborated in \cref{sec:mesh_deform_ex}, thereby driving the Gaussians to represent a caricaturized avatar.

\begin{table}[t]
  \centering
  \setlength{\tabcolsep}{3.5pt} 
  \begin{tabular}{@{}lccccc@{}}
    \toprule
    Method & CLIP-I $\uparrow$ & CLIP-D $\uparrow$ & CLIP-C $\uparrow$ & DINO $\uparrow$ & SD $\uparrow$ \\
    \midrule
    SurFhead     & 0.67 & 0.0006 & 0.944 & 0.757 & 0.460 \\
    Ours & 0.73 & 0.014  & 0.945 & 0.888 & 0.539 \\
    \bottomrule
  \end{tabular}
  \vspace{-6pt}
  \caption{Quantitative comparison for a caricature avatar. Higher is better for all reported metrics.}
  \vspace{-8pt}
  \label{tab:algorithm_results_caricature}
\end{table}

\subsection{Metrics}
\label{subsec:metrics}

Quantitative evaluation of caricature models is inherently challenging due to their under-constrained nature and the lack of ground-truth images.  
We use the following metrics for evaluation:

\begin{itemize}
  \item \textbf{CLIP‐I} (Image–Prompt Similarity) \cite{CLIPscore}: Cosine similarity between the rendered image and text in CLIP space.
  \item \textbf{CLIP‐D} (Directional Similarity)~\cite{gal2021stylegan}: Measures the change between source and edited images against the change between source and edited prompts.
  \item \textbf{CLIP‐C} (Spatial Consistency): Following \cite{Instruct-NeRF2NeRF}, we report CLIP image alignment between adjacent novel views of image embeddings along a novel trajectory.
  \item \textbf{DINO} (Identity/Structure Consistency): Following \cite{zhou_deformable_2024}, we extract DINO \cite{DINO} features from the renders and the corresponding original test frames and compute the cosine similarity of the embeddings.
  \item \textbf{SD} (Score Distillation): Inspired by DreamFusion~\cite{poole_dreamfusion_2022}, we define the reference-free metric as,
   \begin{eqnarray}
    \mathrm{SD}
    = 1 - \frac{1}{BTN}
    \sum_{b,t,n}^{B,T,N}
    \frac{\left\| \epsilon_{\theta}\!\left(x_{t}^{(b,t,n)},\, t\right) - \epsilon_{b,t,n} \right\|_{2}^{2}}
         {\left\| \epsilon_{b,t,n} \right\|_{2}^{2}}.
   \end{eqnarray}

    where \(\epsilon_\theta(x_t,t)\) is the noise predicted by the diffusion model \cite{rombach2022high} at time step \(t\), \(\epsilon\) is the true noise, and \(B, T, N\) refer to the image count, time step, and seed number, respectively.
    Higher SD indicates that the rendered image is more consistent with the training distribution of the diffusion model, which is intended to approximate the natural image distribution.
\end{itemize}
Text prompts are provided in the appendix.
Together, these metrics evaluate: (i) how well the renders reflect the caricature intent (CLIP‐I, CLIP‐D, SD), (ii) identity preservation and the extent to which exaggerations remain localized to caricaturization (DINO, CLIP‐D), and (iii) consistency of generated views across novel trajectories (CLIP‐C).

\subsection{Results}
\label{subsec:results}

\cref{fig:Rendering_Results} presents side-by-side renderings at the target exaggeration level \(\gamma_f\) for our method and the baseline. 
Our approach maintains subject identity while delivering natural, visually pleasing exaggerations that remain consistent across views, and reduces the distortions visible in the baseline.
The figure further illustrates caricature-level controllability by varying \(\gamma\) from \(0\) to \(\gamma_f\), demonstrating continuous control and showing that the approximation in \cref{subsec:Interpolation} successfully supports intermediate exaggeration levels.

For quantitative evaluation, we conduct a comprehensive comparison using the metrics in \cref{subsec:metrics}.
As summarized in \cref{tab:algorithm_results_caricature}, our method consistently surpasses the baseline across all measures, demonstrating that the learned edits faithfully capture the intended caricature while preserving both identity and view-consistency.

\subsection{Diffusion Based Editing}
\begin{figure}[t]
  \centering

  \begin{subfigure}[t]{0.85\linewidth}
    \centering
    \includegraphics[width=\linewidth]{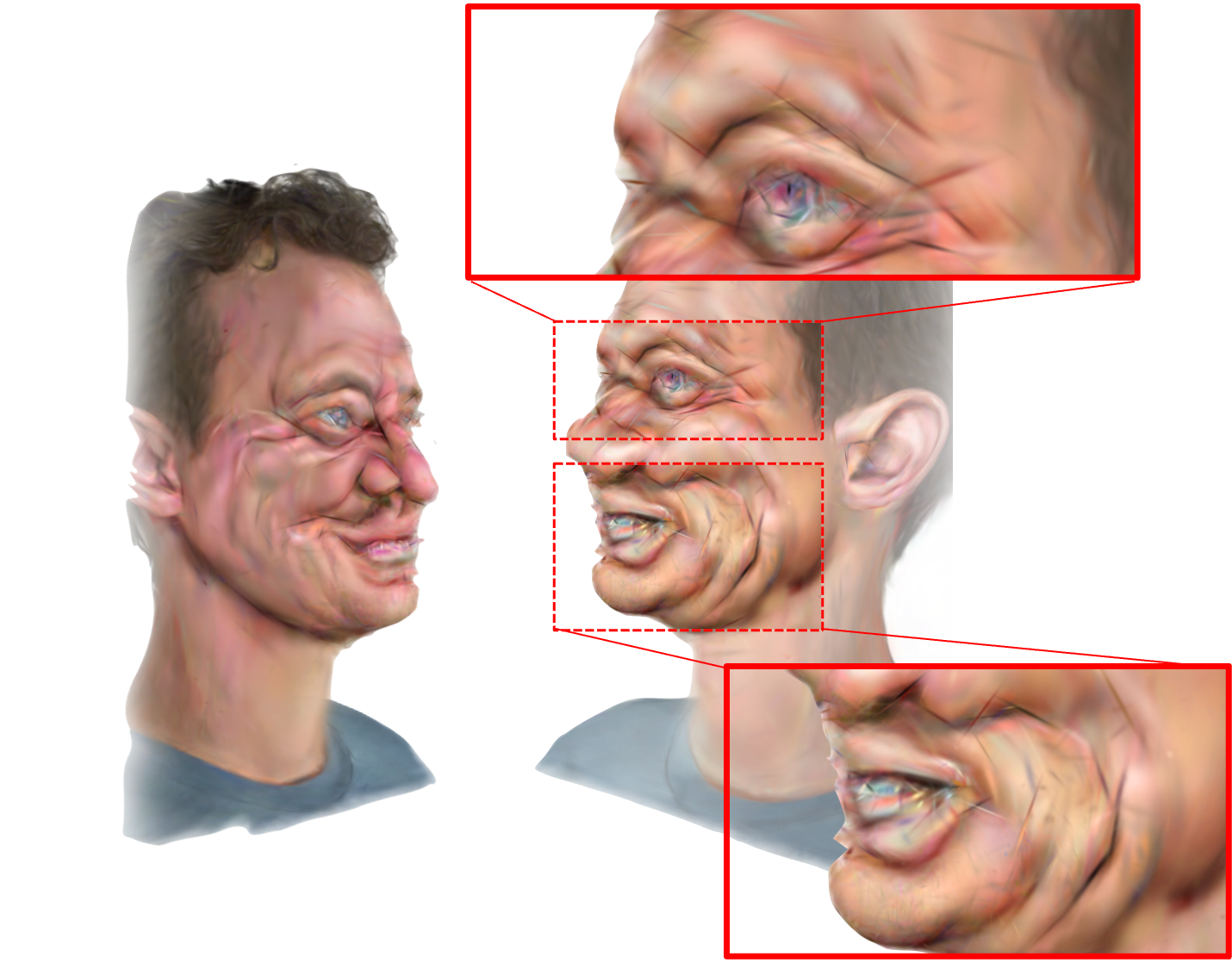}
    \caption{Edit instruction: ``Turn him into a realistic caricature.'' 
    The result exhibits skin-tone shifts and specular degradation.}
    \label{fig:GaussEditor_caricature}
  \end{subfigure}


  \begin{subfigure}[t]{0.85\linewidth}
    \centering
    \includegraphics[width=\linewidth]{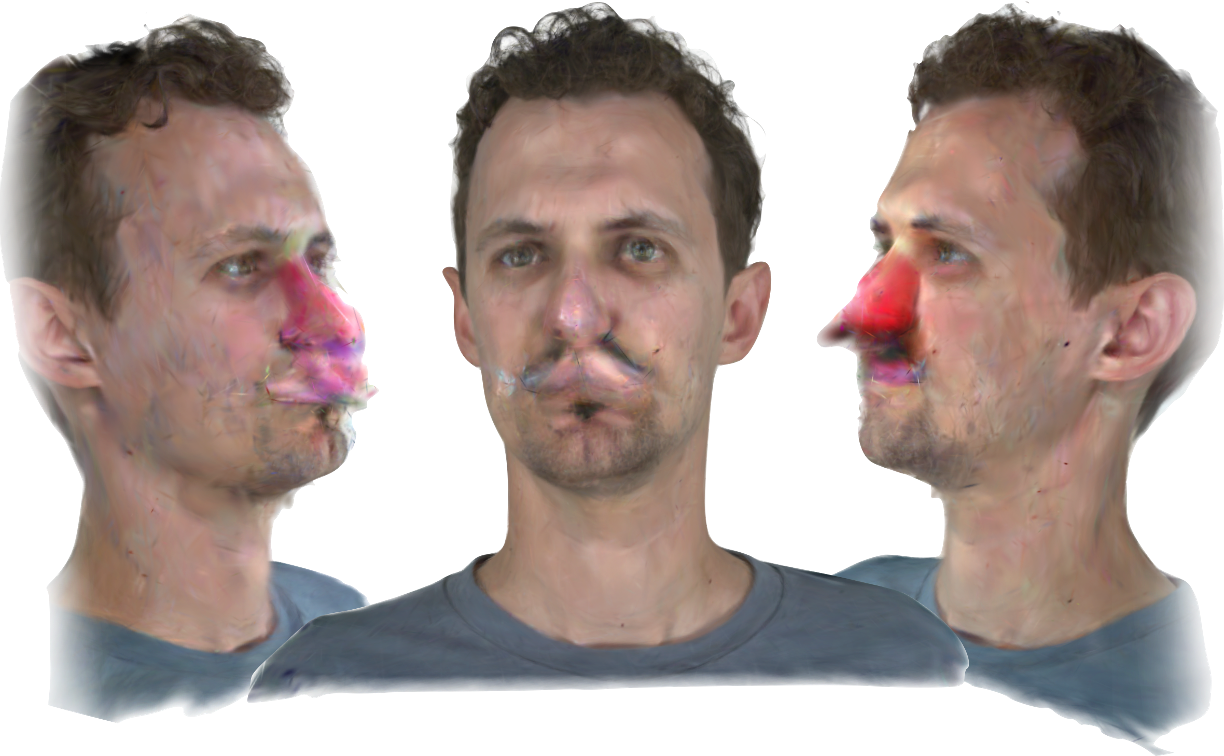}
    \caption{Edit instruction: ``Make his nose bigger.'' 
    The geometry falls apart and color inconsistencies appear across views.}
    \label{fig:GaussEdit_semantic}
  \end{subfigure}
    \vspace{-3pt}
\caption{GaussianEditor~\cite{wang_gaussianeditor_2024} caricaturization attempts. 
  \textbf{(a)} Global edit. \textbf{(b)} Local semantic edit. Both reveal degraded geometry and appearance fidelity, particularly in novel views.
   } 
    \vspace{-10pt}
\label{fig:GaussEditor}
\end{figure}

\vspace{-3pt}
\label{subsec:gaussianEditor}
As an additional baseline, we adapt a diffusion-driven, text-guided, mesh-free 3DGS editor~\cite{chen_gaussianeditor_2024} for caricaturization. 
Using the authors’ implementation, we run 5{,}000 optimization steps per prompt on multiview images of a subject, guided by ControlNet-Pix2Pix.
\cref{fig:GaussEditor_caricature} presents a global edit, while \cref{fig:GaussEdit_semantic} shows a local edit,  manually masked for face and nose, respectively.
While the edits appear visually plausible in individual views, it is evident that, unlike our method, this baseline suffers from (i) geometry drift, (ii) unstable, view-dependent specularities, and (iii) poor multi-view coherence.

\vspace{-3pt}
\section{Ablations}
\vspace{-3pt}
\subsection{Alternated Training}
\vspace{-3pt}
\label{subsec:ablation_supervision}
In this subsection, we demonstrate that training with \agt, generated using LAT, is essential for controlling the caricaturization level.  
As discussed in \cref{subsec:results}, training only on input images fails to generalize: rendering with a caricatured mesh yields heavily degraded outputs.  
In the supplementary, we show that training solely with \agt also fails: neutral renders appear unrealistic, with distorted Gaussian structures.  
These complementary failures underscore the necessity of alternating both forms of supervision for effective caricaturization control.

\subsection{Mask}
\label{subsec:hair_mask}
\begin{figure}[t]
  \centering
  \includegraphics[width=0.95\linewidth]{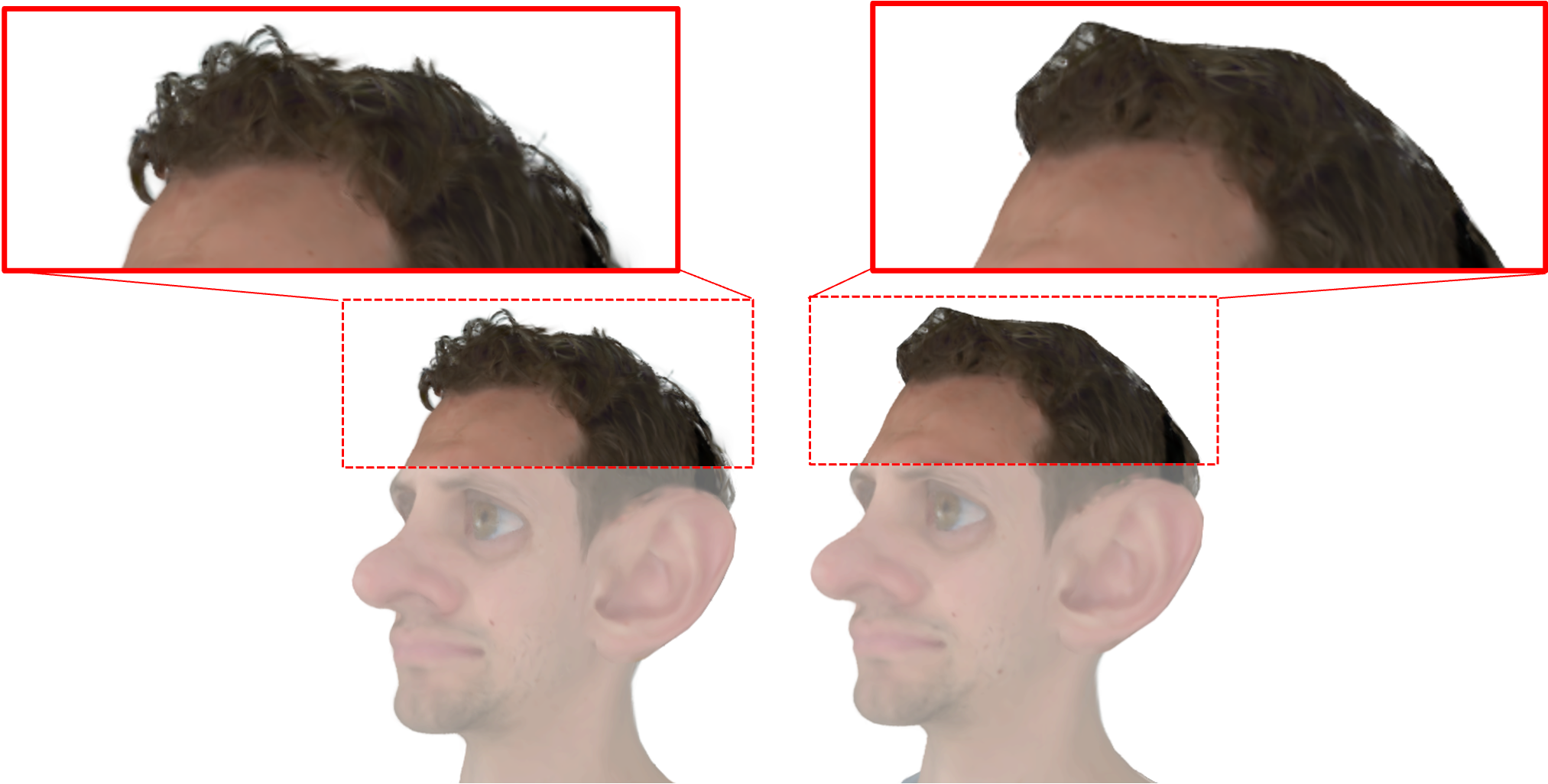}
    \vspace{-3pt}
  \caption{Ablation on hair masking. Without masking, \agt introduces visible artifacts in hair regions. 
  Masking and freezing Gaussians associated with hair during \agt supervision effectively prevents these artifacts.}
  \vspace{-9pt}
  \label{fig:hair_error}
\end{figure}

Due to the nature of \agt generation, certain fine details, most notably hair, are often misrepresented during the caricature stage. 
To address this, we identify hair regions of the mesh and freeze the corresponding Gaussian parameters with a suitable mask during \agt supervision iterations, thereby preventing updates in those regions when the caricature is rendered (see \cref{subsec:LAT_GT}). 
\cref{fig:hair_error} illustrates the effect: on the left, hair regions are masked and remain frozen, whereas on the right they are unfrozen and allowed to train freely, resulting in unnaturally plastic-looking hair.

\section{Limitations}
\label{sec:limitations}

While our method provides a powerful framework for photorealistic 3D caricaturization, several limitations remain. 
Although our approach improves upon the baseline, residual specularity artifacts persist, and small eyelid inaccuracies—amplified by over-stretching in LAT, become visually noticeable.  
This effect also extends to hair: training Caricature 3DGS hair with input-view supervision alone (without \agt) substantially alleviates the issue. However, in some cases, we observe slight over-smoothing of the hair. 
Qualitative examples of these effects are provided in the supplementary material.  
Finally, the deformed FLAME mesh does not fully span the space of facial expressions. For instance, eyelid closure in caricatured results is imperfect: eyes that should be completely shut under certain expressions often remain slightly open, leading to misrepresentations of eyelid geometry in the final caricature.

\vspace{-2pt}
\section{Discussion}
\vspace{-2pt}
This work demonstrates that curvature-driven geometric deformation and mesh-rigged 3D Gaussian Splatting (3DGS) can be combined into a single, controllable avatar model that remains photorealistic under large exaggerations.
The key is a training scheme that alternates supervision between real views and generated pseudo–ground-truth caricature views, produced using per-triangle Local Affine Transformations (LAT) with reliability masks.
One Gaussian set is capable of jointly learning both natural and caricatured appearance while retaining identity and expression. 
Prior work indicates that deliberate shape exaggeration can amplify discriminative geometric cues for recognition \cite{sela_computational_2015}.
Looking ahead, we hypothesize that integrating our controllable exaggeration as a plug-in augmentation within face-recognition pipelines could improve robustness to pose and expression variability.
Finally, coupling our geometry-grounded deformations with diffusion-based editors may enable semantically guided edits that are both photorealistic and extend beyond appearance-only changes to joint control of shape and appearance.
\label{sec:future}

{\small
\bibliographystyle{ieeenat_fullname}
\bibliography{main}
}
\clearpage
\setcounter{page}{1}
\maketitlesupplementary

\section{Implementation considerations}
\label{subsec:implementation_appendix}

Unless stated otherwise, we optimize each subject’s 3D Gaussian Splatting model for \(120{,}000\) iterations, adhering to SurFhead’s training protocol and evaluation split~\cite{lee_surfhead_2024}. All experiments are run on a single NVIDIA RTX~3090 (24\,GB VRAM). The optimization time per subject is \(\approx 4\) hours (this is offline training time, not rendering runtime.)

We used the NeRSemble dataset~\cite{face2face} with 10 subjects, 4 emotions (EMO), and 6 expressions (EXP). Expression \texttt{EXP2} is held for testing and Camera~8 serves as the validation view during training.

Caricaturization is performed once at the beginning of the training by solving the unconstrained Poisson equation, deforming the FLAME base template with \(\gamma=0.25\) (\(\approx\!1\,\mathrm{min}\)). 

Because FLAME uses a shared template across subjects, the deformed surface is saved and reused for all subjects.
Unless stated otherwise, we report metrics over 256 frames from the rendered test sequence, aggregated across all camera viewpoints.

\paragraph{CLIP configuration.}
For text–image alignment, we use OpenAI CLIP with the \texttt{ViT-B/32} backbone and the library’s default preprocessing.

Prompts are:
\textbf{Source:} ``A realistic neutral head with natural lighting.''\;
\textbf{Edit:} ``A photorealistic caricature of a head with a highly exaggerated nose and large ears, under natural lighting.''

\paragraph{Defaults inherited.}
The optimizer, learning rate schedule, degree of spherical harmonics, and Gaussian growth/pruning follow the SurFhead~\cite{lee_surfhead_2024} configuration unless otherwise specified.

\section{Linear Model and Error Analysis}
\label{subsec:secant-analysis-append}

\paragraph{Notation.}
Let \(S(u,v)\) be a parametric surface, where \((u,v)\in\mathbb{R}^2\), with a metric \(G\) and \(K\) denotes the Gaussian curvature at each point of the surface \(S\), and
\begin{eqnarray}
w(\gamma) &=& |K|^{\gamma} \;=\; e^{\gamma L}, \qquad L\; \equiv\;  \ln|K|. \label{weights}
\end{eqnarray}
For \(\gamma\in[0,\gamma_f]\), denote by \(S_\gamma\) the solution of the weighted Poisson problem with Dirichlet boundary condition \(x^*\) on \(\partial S\).

To avoid degeneracies at \(K=0\), we use \(\epsilon\) to stabilize the magnitude.
Note, for convenience we refer to \(\) as \(|K|_\epsilon=\sqrt{K^2+\epsilon^2}\) with fixed \(\epsilon>0\). For brevity we write \(|K|\) to denote this stabilized quantities. 

\medskip
\noindent\textbf{1) Poisson equation with secant weights.}
The original family is defined by
\begin{eqnarray}
\Delta_G S_\gamma &=&
\nabla_G\!\cdot\!\big(w(\gamma)\,\nabla_G S\big).
\end{eqnarray}
Note, that \(S_0\) and \(S_{\gamma_f}\) refer to \(\gamma=0\) and \(\gamma=\gamma_f\), respectively. Define the vertex blend,
\begin{eqnarray}
\label{eq:linear_gamma}
S_{\mathrm{blend}}(\gamma) &=& (1-\alpha)\,S_0 \;+\; \alpha\,S_{\gamma_f}, \quad \alpha\; \equiv \;\frac{\gamma}{\gamma_f}.
\end{eqnarray}
 
By linearity of \(\Delta_G\) and Equation \eqref{eq:linear_gamma}
\begin{eqnarray}
\Delta_G S_{\mathrm{blend}}(\gamma)
&=& (1-\alpha)\,\Delta_G S_0 \;+\; \alpha\,\Delta_G S_{\gamma_f} \cr
&=& \nabla_G\!\cdot\!\Big(w_{\mathrm{sec}}(\gamma)\,\nabla_G S\Big),
\end{eqnarray}
where \emph{secant weight} is
\begin{eqnarray}
w_{\mathrm{sec}}(\gamma) = 1 + \frac{\gamma}{\gamma_f}\big(|K|^{\gamma_f}-1\big).
\end{eqnarray}
Thus \(S_{\mathrm{blend}}(\gamma)\) solves the exact Poisson equation at level \(\gamma\) with \(w(\gamma)\) replaced by \(w_{\mathrm{sec}}(\gamma)\), and \(S_{\mathrm{interp}}\rvert_{\partial S}=x^*\) (see \eqref{eq:constrained-ls} for \(x^*\)).

\medskip
\noindent\textbf{2) Remainder and properties}
The secant \(w_{\mathrm{sec}}\) is the linear interpolant of \(w\) in \([0,\gamma_f]\). 
By the classical interpolation remainder for \(C^2\) functions on a closed interval (e.g., \cite[Thm.~3.1]{BurdenFaires2010}, \cite[\S3.3]{AtkinsonHan2009}), for every \(\gamma\in[0,\gamma_f]\) there exists \(\xi(\gamma)\in(0,\gamma_f)\) such that
\begin{eqnarray}
w_{\mathrm{sec}}(\gamma)-w(\gamma) &=& \frac{w''(\xi)}{2}\,\gamma(\gamma_f-\gamma).
\end{eqnarray}
Since \(w''(\gamma)=L^2 e^{\gamma L}\), we get
\begin{eqnarray}
w_{\mathrm{sec}}(\gamma)-w(\gamma) &=& \frac{L^2}{2}\,e^{\xi L}\,\gamma(\gamma_f-\gamma).
\end{eqnarray}
The secant model is exact at both endpoints (where $\alpha=0$ and $\alpha =1$, yielding a analytic expression in \([0,\gamma_f]\)  preserving the convexity-induced non-negativity.


\noindent Since \(w''\ge 0\), \(\gamma\mapsto w(\gamma)\) is convex, hence

\(w_{\mathrm{sec}}-w\) is nonnegative on \([0,\gamma_f]\) and vanishes at the endpoints.

In particular, at \(\gamma=\gamma_f/2\),
\begin{eqnarray}
\big|w_{\mathrm{sec}}(\tfrac{\gamma_f}{2})-w(\tfrac{\gamma_f}{2})\big|
&\le& \frac{\gamma_f^2}{8}\,L^2\,\max\ (1,e^{\gamma_f L}\ ).
\label{eq:bound_appendix}
\end{eqnarray}
The maximum of this \emph{upper bound} occurs at \(\gamma_f/2\) because \(\gamma(\gamma_f-\gamma)\) is maximized there.

\medskip
\noindent\textbf{3) Poincar\'e and Lax--Milgram for residual bound.}

Throughout, we approximate the $\gamma$–dependent weight $w(\gamma)=|K|^{\gamma}$ by its
secant $w_{\mathrm{sec}}(\gamma)$ to enable a cheap vertex blend
instead of solving a new Poisson problem for each $\gamma$.
To justify this alternative, we should \emph{quantify} how the
weight error propagates to a \emph{geometric residual}
\(
\delta S(\gamma)\equiv S(\gamma)-S_{\mathrm{blend}}(\gamma).
\)
The goal here is to derive a norm bound on $\delta S$ that depends only on:
(i) ellipticity and Poincar\'e constants of the domain,
(ii) the magnitude of $\nabla_G S_0$, and
(iii) the scalar secant remainder from Appendix~\cref{eq:bound_appendix}.
This yields a mesh and metric agnostic error budget for the blend.

\paragraph{Setting (frozen operator).}
Let \((S,G)\) be a compact Riemannian surface with Lipschitz boundary \(\partial S\).
We impose Dirichlet conditions \(u\big|_{\partial S}=0\).

We fix the differential operators on the surface \(S\),
namely, the gradient and the divergence w.r.t metric \(G\).

Let $V \equiv H^1_0(S)$ and define
\begin{eqnarray}
a(u,v) &=& \int_S \langle \nabla_G u,\nabla_G v\rangle_G\,dA_G \cr
\|u\|_V &\equiv& \|\nabla_G u\|_{L^2(S)}. 
\end{eqnarray}


\noindent We also define the \emph{dual norm} by 
\begin{eqnarray}
\|F\|_{V'} \;\equiv \; \sup_{v\in V\setminus\{0\}} \frac{|F(v)|}{\|v\|_V}. 
\label{eq:supremum_dual}
\end{eqnarray}

\noindent Using \emph{Poincar\'e inequality}, there exists $C_P>0$ such that, for all \(u\in H^1_0(S)\),
\begin{eqnarray}
\|u\|_{L^2(S)} \;\le\; C_P\,\|\nabla_G u\|_{L^2(S)} \;=\; C_P\,\|u\|_V. 
\end{eqnarray}
Hence $\|u\|_V$ is a true norm on $H^1_0(S)$ and is equivalent to the standard $H^1$-norm on $H^1_0(S)$.

\noindent By Cauchy--Schwarz,
\begin{eqnarray}
|a(u,v)| &\le& \|u\|_V\,\|v\|_V \quad\text{(boundedness)}, \cr
a(v,v) &=& \|v\|_V^2 \quad\;\;\;\text{(coercivity with }\alpha=1\text{)} 
\label{eq:coercivity}
\end{eqnarray}
where coercivity means that there exists $\alpha>0$ such that
\[
a(v,v)\;\ge\;\alpha\,\|v\|_V^2\quad\forall\,v\in V.
\]

\noindent\textbf{Lax--Milgram.} 
If $a$ is bounded and coercive on the Hilbert space $V$ and $F\in V'$ is bounded, then, there exists a unique solution
$u\in V$, solving $a(u,v)=F(v)$ for all $v\in V$, with estimate
\begin{eqnarray}
\|u\|_V \;\le\; \frac{1}{\alpha}\,\|F\|_{V'} \;\overset{\eqref{eq:coercivity}}{=}\; \|F\|_{V'} .
\end{eqnarray}


For each $\gamma$, we solve the weighted Poisson PDE given by 
\begin{eqnarray}
\Delta_G S_\gamma & = & \operatorname{\nabla_G}\!\big(w(\gamma)\,\nabla_G S\big), \qquad S_\gamma\!\big|_{\partial S}=x^*. 
\end{eqnarray}
Let $S_{\mathrm{blend}}(\gamma)=(1-\alpha)S_0+\alpha S_{\gamma_f}$ with $\alpha=\gamma/\gamma_f$, and define
\begin{eqnarray}
\psi(\gamma) &\equiv& w_{\mathrm{sec}}(\gamma)-w(\gamma) \cr
\mathcal{R}_\Delta(\gamma) & \equiv & \operatorname{\nabla_G}\!\big(\psi\,\nabla_G S\big). 
\end{eqnarray}
Define $F\in V'$ (weak residual functional) by
\begin{eqnarray}
F(v) &=& \langle \mathcal{R}_\Delta, v\rangle \cr
&= &\int_S \big(\operatorname{\nabla_G}(\psi\,\nabla_G S)\big)\,v\,dA_G \cr
&=& -\int_S \psi\,\langle \nabla_G S,\nabla_G v\rangle_G\,dA_G,
\end{eqnarray}
with $v\big|_{\partial S}=0$.

Using the dual norm and by Cauchy--Schwarz and $\|\psi\|_{L^\infty}$-bound, we readily have
\begin{eqnarray}
|F(v)| &\le& \|\psi\|_{L^\infty(S)}\,\|\nabla_G S\|_{L^2(S)}\,\|\nabla_G v\|_{L^2(S)} \cr 
&=& \|\psi\|_{L^\infty}\,\|\nabla_G S\|_{L^2(S)}\,\|v\|_V, 
\end{eqnarray}
and using \eqref{eq:supremum_dual} we get
\begin{eqnarray}
\|F\|_{V'}
&\le& \|\psi\|_{L^\infty}\,\|\nabla_G S\|_{L^2(S)}. 
\end{eqnarray}

\noindent Let $\delta S \equiv S_{\mathrm{blend}}-S_\gamma$. 
Subtract the weak forms for $S_{\mathrm{blend}}$ and $S_\gamma$ to obtain
\begin{eqnarray}
a(\delta S, v) &=& a(S_{\mathrm{blend}}, v) - a(S_\gamma, v) \cr
&=& \int_S w_{\mathrm{sec}}\,\langle \nabla_G S,\nabla_G v\rangle_G\,dA_G \cr
     & & - \int_S w(\gamma)\,\langle \nabla_G S,\nabla_G v\rangle_G\,dA_G \cr
&=& \int_S \psi\,\langle \nabla_G S,\nabla_G v\rangle_G\,dA_G \cr
&=& -\int_S \operatorname{\nabla_G}\!\big(\psi\,\nabla_G S\big)\,v\,dA_G 
\quad (*) \cr
&\equiv& -\,F(v). 
\end{eqnarray}

\noindent Where in (*) we use integration by parts and Dirichlet boundary conditions on \(\partial S\).

Testing with $v=\delta S$ and using coercivity and duality,
\begin{eqnarray}
\|\delta S\|_V^2 &=& a(\delta S,\delta S) \cr
&=& -\,F(\delta S) 
\;\le \;\|F\|_{V'}\,\|\delta S\|_V \cr
\Rightarrow
\|\delta S\|_V &\le& \|F\|_{V'}. 
\end{eqnarray}
Combining with the bound on $\|F\|_{V'}$ yields the \emph{energy} estimate
\begin{eqnarray}
\|\delta S\|_V &\le& \|\psi\|_{L^\infty(S)}\,\|\nabla_G S\|_{L^2(S)} \cr \cr
\|\delta S\|_V &\le& \|w_{\mathrm{sec}}-w\|_{L^\infty}\,\|\nabla_G S\|_{L^2(S)}. 
\end{eqnarray}

\paragraph{Optional $L^2$ bound.}
By Poincar\'e on $H^1_0(S)$,
\begin{eqnarray}
\|\delta S\|_{L^2(S)} &\le& C_P\,\|\delta S\|_V \cr
&\le&  C_P\,\|w_{\mathrm{sec}}-w\|_{L^\infty}\,\|\nabla_G S\|_{L^2(S)}. 
\end{eqnarray}

In summary, the secant error bound yields the energy bound for the residual \(\delta S\) by

\begin{equation}
\begin{aligned}
\|\delta S(\gamma)\|_{L^2} &\lesssim &
C_P(\ln\|K\|)^2 e^{\max(0,\gamma_f\ln\|K\|)} \cr
&\, &\times \gamma(\gamma_f-\gamma)\,\|\nabla_G S\|_{L^2(S)}.\label{eq:final_derivation}
\end{aligned}
\end{equation}
which depends on geometric constants of the domain ($C_P$).
The curvature in \eqref{eq:final_derivation} is evaluated at its global maximum
\begin{equation}
\|K\| = K_\infty = \max_{s\in S}\,|K(s)|
\end{equation}

We note that \(S_0 = S\) (for \(\gamma=0\) by definition since there is no deformation done to \(S\)), hence \eqref{eq:final_derivation} can be written using either terms.

\section{Caricature \agt via one-shot stylization}
\label{sec:oneshot_agt}
\begin{figure}[t]
  \centering

  \begin{subfigure}[t]{\linewidth}
    \centering
    \includegraphics[width=\linewidth]{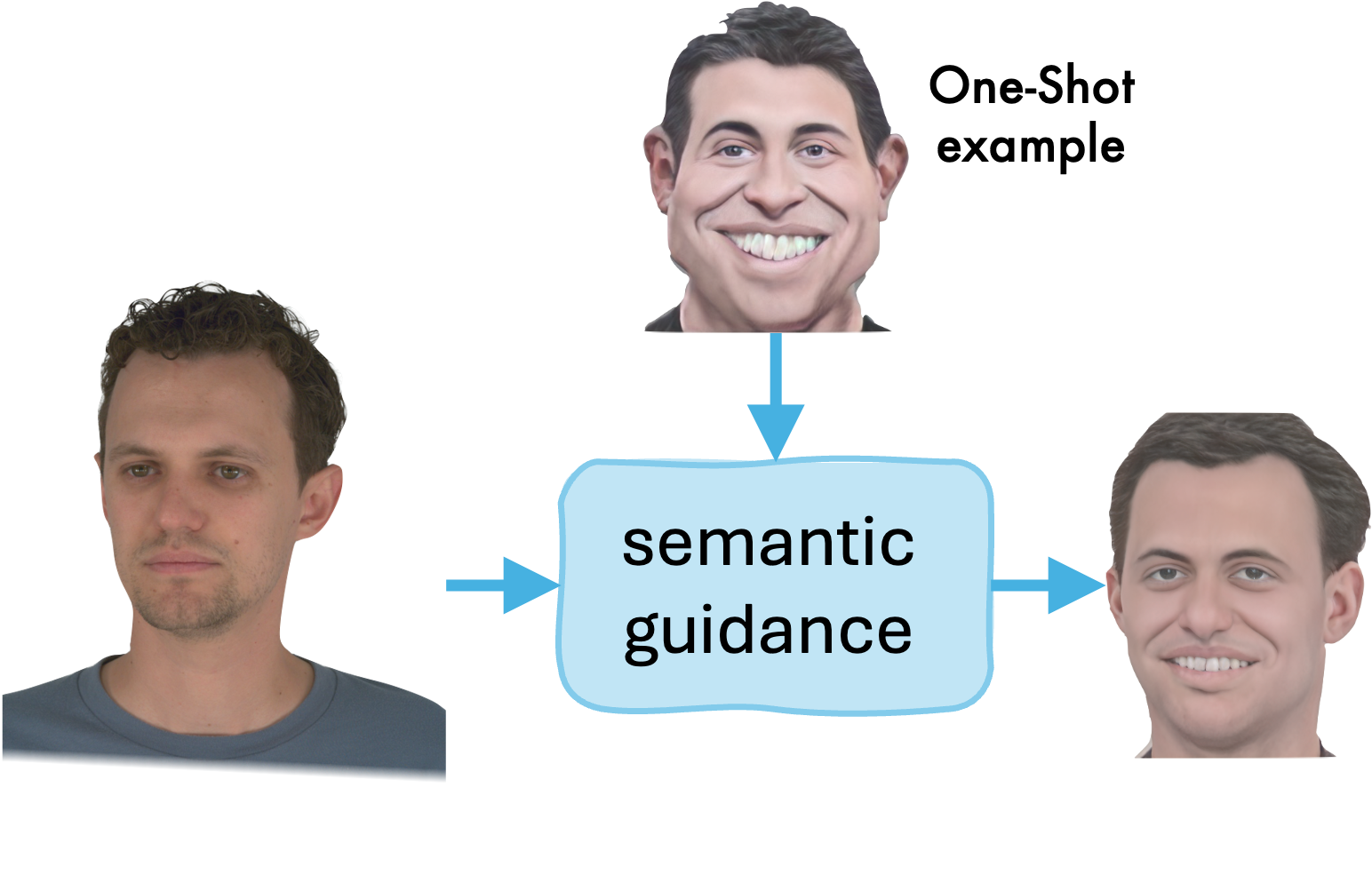}
    \caption{Deformable StyleGAN~\cite{zhou_deformable_2024}: stylization conditioned on a target style exemplar.}
    \label{fig:doesfs:a}
  \end{subfigure}

  \vspace{3pt}

  \begin{subfigure}[t]{\linewidth}
    \centering
    \includegraphics[width=\linewidth]{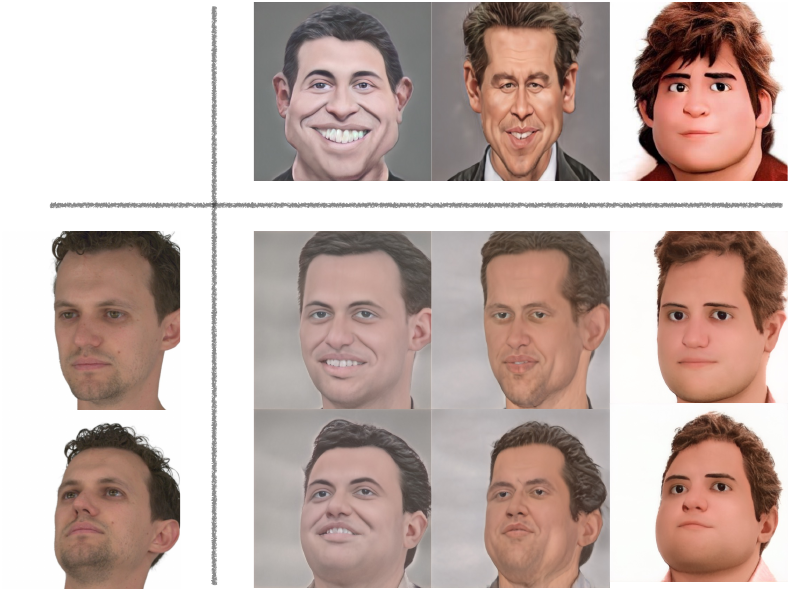}
    \caption{View variation induces identity drift and structural artifacts (\eg neck geometry).}
    \label{fig:doesfs:b}
  \end{subfigure}

  \vspace{3pt}

  \begin{subfigure}[t]{\linewidth}
    \centering
    \includegraphics[width=\linewidth]{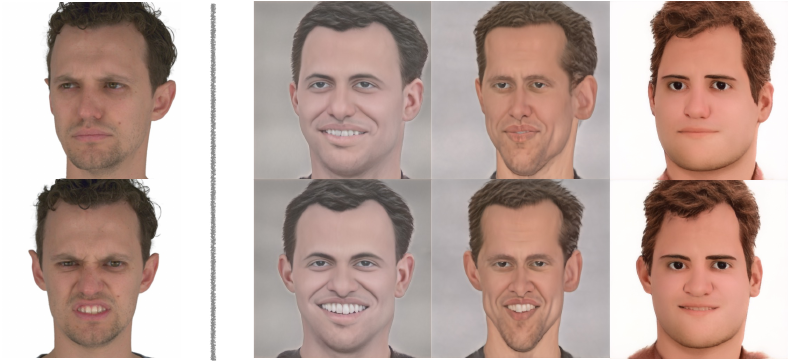}
    \caption{Expressions are not preserved, outputs bias toward the style exemplar (\eg persistent smile, forward gaze).}
    \label{fig:doesfs:c}
  \end{subfigure}

  \caption{Limitations of one-shot stylization for caricature. Identity–expression entanglement and lack of view/expression consistency hinder 3DGS supervision.}
  \label{fig:doesfs_ablation}
\end{figure}

As discussed in \cref{sec:method}, one-shot stylization methods (e.g., Deformable StyleGAN~\cite{zhou_deformable_2024}) address the natural-caricature domain gap by aligning DINO features and adapting a pretrained GAN to a single caricature exemplar. 
Given a target style image (\cref{fig:doesfs:a}), they synthesize stylized outputs for arbitrary inputs. In practice, we observe pronounced identity–expression entanglement, which degrades both identity fidelity and expression accuracy (\cref{fig:doesfs_ablation}). Moreover, the outputs are not consistent across viewpoints or expressions: under view changes or when transferring expressions from the source, the method exhibits structural drift and a collapse toward the reference style (\cref{fig:doesfs:b,fig:doesfs:c}), limiting its suitability for our 3DGS reconstruction setting.

\paragraph{Protocol.}
We ran \cite{zhou_deformable_2024} using the official implementation, employing \texttt{Style1}, \texttt{Style2}, and \texttt{Style3} as target style exemplars and \texttt{EMO3}, \texttt{EMO4} for expression prompts.

\begin{figure}[!htbp]
  \centering
  \includegraphics[width=0.9\linewidth]{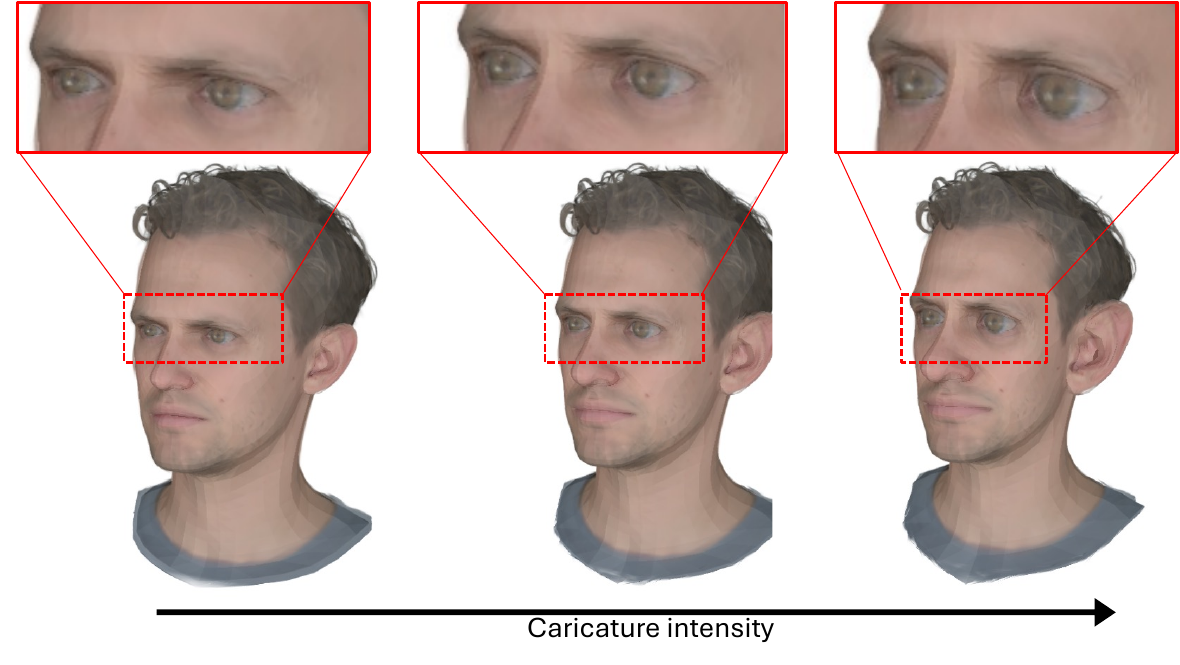}
  \caption{FLAME–image misregistration under increasing caricature strength~$\gamma$. Projection drift concentrates on thin, high-curvature structures (eyelids/iris rim) and grows with \(\gamma\), introducing erroneous supervision if used unfiltered.}
  \label{fig:eyelids_ablation}
\end{figure}

\section{Masking and \agt}
\label{sec:masking_and_latgt}
As noted in \cref{subsec:LAT_GT}, \agt\ supervision is constructed by projecting the FLAME mesh, fitted to each original frame, onto the image. Consequently, the quality of \agt\ inherits any mesh–image misregistration. In practice, small fitting errors that are negligible at \(\gamma{=}0\) are amplified as the caricature strength increases, with the most visible drift around delicate geometry such as the eyelids and eyeballs; see \cref{fig:eyelids_ablation}. In addition, the deformation can reveal triangles that were occluded in the original projection (e.g., along the eyelid crease), creating pixels with no reliable photometric support.

To prevent these failure modes, we build a visibility-aware \agt\ mask. We (i) suppress supervision on triangles that become newly visible at nonzero \(\gamma\) relative to the original projection, and (ii) mask anatomically fragile regions prone to amplified alignment error (eyelids, ear tips). This filtering removes inconsistent labels before they reach Gaussians anchored to those areas, yielding cleaner gradients and more stable appearance/geometry during training. The resulting \agt\ thus preserves the benefits of deformation-aware supervision while avoiding artifacts introduced by projection drift and occlusions.

\section{Ablation: Alternating Supervision}
\label{subsec:ablation_alternation}

\paragraph{Setup.}
As motivated in \cref{subsec:ablation_supervision}, we seek a \emph{single} 3DGS model that renders both the original avatar ($\gamma{=}0$) and its caricatured counterpart ($\gamma{=}\gamma_f$). We compare three training schedules using identical budgets: 
(i) \emph{Original-only}: supervision from original frames only.
(ii) \emph{\agt-only}: supervision from caricatured (\agt) frames only.
(iii) \emph{Alternating (ours)}: alternating mini-batches from both sources. 
We set the target exaggeration to $\gamma_f{=}0.25$ and evaluate along the interpolation path $\gamma \in \{0, 0.10, 0.15, 0.20, 0.25\}$.

\begin{figure*}[t]
  \centering
  \makebox[\textwidth][c]{%
    \includegraphics[width=1\textwidth]{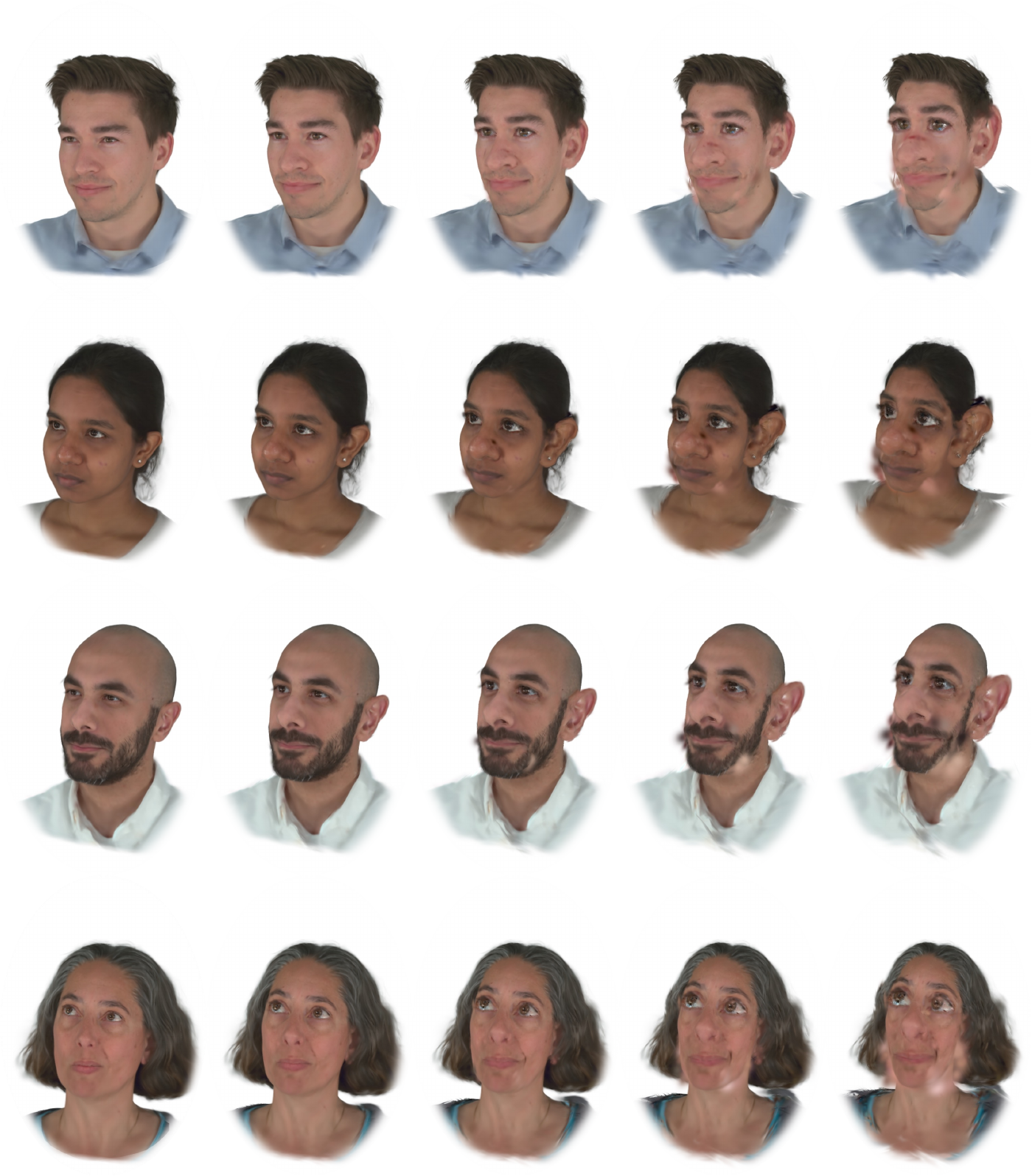}%
  }
  \caption{
  {Training on original frames only}
    }
    \vspace{-6pt}
  \label{fig:original_interpolation_ablation}
\end{figure*}

\begin{figure*}[t]
  \centering
  \makebox[\textwidth][c]{%
    \includegraphics[width=1\textwidth]{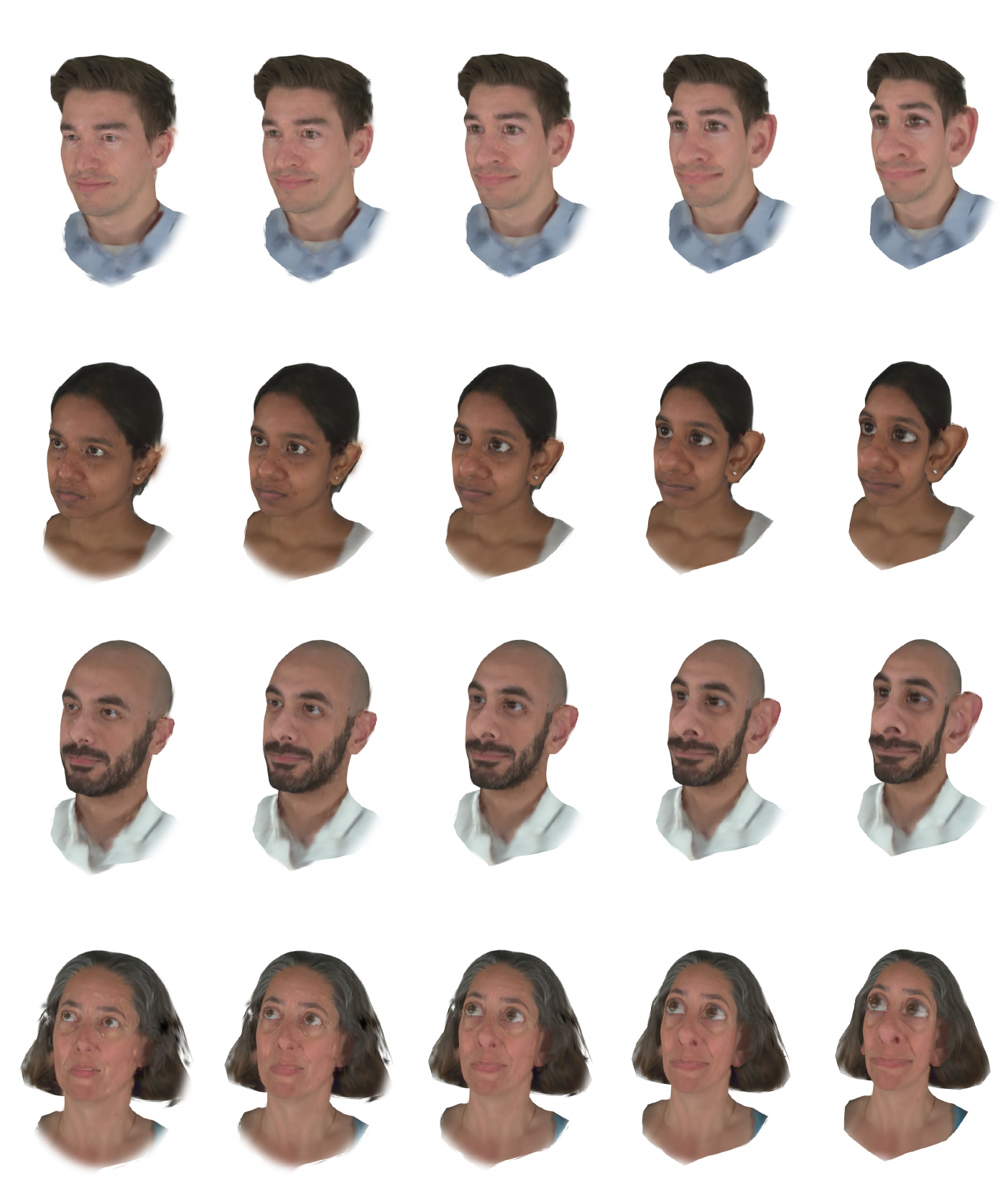}%
  }
  \caption{
  {Training on \agt frames only.}
    }
    \vspace{-6pt}
  \label{fig:caricature_holes_ablation}
\end{figure*}

\paragraph{Findings.}
Original-only (i) fits the undeformed scene well but fails to generalize to caricatured geometry \cref{fig:original_interpolation_ablation}, yielding visible distortions under nonzero $\gamma$. 
Conversely, \agt-only (ii) represents the caricatured avatar but degrades markedly at $\gamma{=}0$. In addition, \agt-only exhibits systematic artifacts around hair and other structures that extend beyond the tracked mesh support (\eg holes or under-coverage), because those pixels are never directly supervised in the warped domain, see \cref{fig:caricature_holes_ablation}. 

\begin{figure*}[t]
  \centering
  \makebox[\textwidth][c]{%
    \includegraphics[width=1\textwidth]{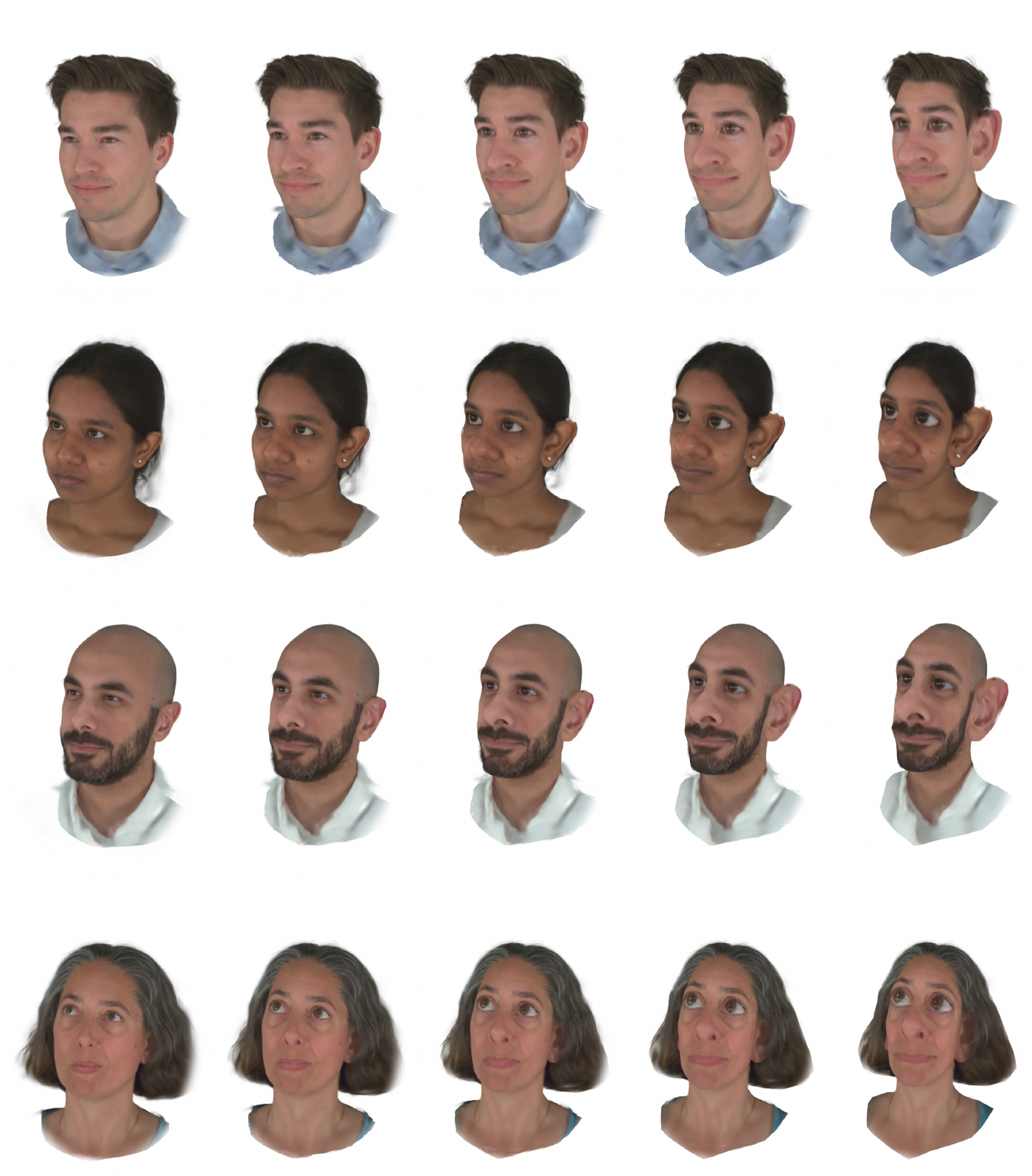}%
  }
  \caption{
  {Training on both original and \agt frames interleaved}
    }
    \vspace{-6pt}
  \label{fig:ours_interpolation_ablation}
\end{figure*}

Our alternate schedule (iii) maintains high fidelity at both endpoints and produces smooth interpolation across $\gamma$ (see \cref{fig:ours_interpolation_ablation}), avoiding the hair/occlusion failures seen in (ii). Practically, alternating acts as a simple multi-domain regularizer, as it preserves appearance outside the mesh support (from original frames) while learning the exaggerated geometry and view-dependent effects required by \agt.

\paragraph{Conclusions.}
Alternating supervision is necessary to obtain a \emph{single} 3DGS that is faithful at $\gamma{=}0$ and $\gamma{=}\gamma_f$ and stable along the interpolation path, while training on either domain alone leads to domain-specific overfitting and characteristic failure modes.

\end{document}